\documentclass[manuscript,screen]{acmart}
\AtBeginDocument{%
  }

 \usepackage[framemethod=tikz]{mdframed}
\usepackage{tikz}
\usetikzlibrary{mindmap}
\usepackage{amsmath}

\usepackage{filecontents}
\usepackage{tcolorbox}
\usepackage{multirow}
\usepackage{subcaption}
\usepackage{hyperref}
\usepackage{supertabular}
\usepackage[textsize=tiny]{todonotes}
\usepackage{graphicx}
\usepackage[threshold=100,thresholdtype=words,autopunct=false]{csquotes}

\setcopyright{acmlicensed}
\copyrightyear{2026}
\acmYear{2026}
\acmDOI{XXXXXXX.XXXXXXX}
\acmISBN{978-1-4503-XXXX-X/2026/06}

\begin{document}

\title{Advancing Inclusivity in Cybersecurity Education: Integrating Intersectionality to Enhance Student Engagement in Australian Higher Education Curriculums — Strategies, Barriers, and Future Directions}


\author{Nalin A. G. Arachchilage}
 \authornote{Both authors contributed equally to this research.}
 \email{nalin.arachchilage@rmit.edu.au}
 \orcid{1234-5678-9012}
 \author{Asangi Jayatilaka}
 \authornotemark[1]
 \email{asangi,jayatilaka@rmit.edu.au}
 \affiliation{
  \institution{School of Computing Technologies, RMIT}
  \city{Melbourne}
  \state{Victoria}
  \country{Australia}
 }

 \author{Senuri Wijenayake }
 \affiliation{%
  \institution{School of Computing Technologies, RMIT}
   \city{Melbourne}
   \country{Australia}}
 \email{senuri.wijenayake@rmit.edu.au}

 \author{David Herbert}
 \affiliation{%
  \institution{ University of Tasmania}
  \city{Hobart}
  \country{Australia}
 }
 \email{DAVID.HERBERT@UTAS.EDU.AU}

 \author{Kaie Maennel}
  \affiliation{%
  \institution{AdelaideUniversity}
  \city{Adelaide}
  \state{South Australia}
  \country{Australia}}
 \email{ kaie.maennel@adelaide.edu.au}

 \author{Nicole Herbert}
  \affiliation{%
  \institution{ University of Tasmania}
  \city{Hobart}
  \country{Australia}
 } \email{nicole.herbert@utas.edu.au}

 \author{Claudia Szabo}
 \affiliation{%
  \institution{Adelaide University}
  \city{Adelaide}
  \state{South Australia}
  \country{Australia}}
  \email{claudia.szabo@adelaide.edu.au}

 \author{Gabrielle Murray}
 \affiliation{%
  \institution{Office of Indigenous Education Research and Engagement, RMIT University}
   \city{Melbourne}
   \country{Victoria}}
 \email{gabrielle.murray@rmit.edu.au}

 \author{Gary Thomas}
 \affiliation{%
  \institution{Office of Indigenous Education Research and Engagement, RMIT University}
   \city{Melbourne}
   \country{Victoria}}
\email{gary.thomas@rmit.edu.au}



\begin{abstract}
 Australian women, gender-diverse individuals, and culturally and linguistically diverse (CALD) communities
are often more susceptible to phishing and other forms of cybercrimes due to factors such as language barriers,
limited access to cybersecurity education, and social isolation. These communities encounter substantial
obstacles both entering and progressing in the cybersecurity field. In Australia, the Higher Education sector
still leans heavily on a largely uniform cybersecurity curriculum, focusing heavily on technical proficiency,
overlooking the vital impact of intersectionality and user-centered thinking for boosting student engagement
and learning. Without gender inclusivity and proper consideration of intersectionality forms such as CALD, the
workforce is deprived of the varied perspectives necessary to tackle today’s intricate cybersecurity issues. In
this study, we conducted semi-structured interviews with 15 experienced academics teaching and coordinating
cyber security programs from a diverse range of Australian universities, covering all states, to explore their
perspectives on: i) current strategies for addressing the women, gender-diverse and CALD perspective in
cyber security education in the Australian HE sector; ii) barriers to incorporate women, gender-diverse and
CALD perspective in cybersecurity curriculums in higher education; iii) future work and support that is
needed. Our research highlights a lack of systematic methods for integrating intersectional perspectives into
cybersecurity curriculums. In particular, we identified four key barriers and four areas where support and
future efforts are needed to address this issue. Our findings offer vital insights that can substantially guide
curriculum development in cybersecurity education. Additionally, these insights create a comprehensive
roadmap for academic institutions, government bodies, accreditation bodies, and industry leaders to cultivate
a more inclusive learning environment.
\end{abstract}



\maketitle
\section{Introduction}
Cybersecurity in the digital age has evolved into a sophisticated battleground where the most vulnerable targets are not systems, but people. Phishing attacks, social engineering tactics, and other forms of cyber exploitation thrive on human error, exploiting individuals as the weakest link in information security \cite{desolda2021human, zhuo2023sok} - phishing propagates other cyber-attacks such as ransomware or malware. The Australian Cyber Security Strategy 2023–2030 emphasizes that a critical skill gap exists in addressing these human vulnerabilities (via social engineering like phishing scams), with cybercriminals increasingly honing attacks to exploit personal, cultural, and situational weaknesses \cite{farlowcase}. This gap is especially pronounced when we consider intersectionality \cite{thomas2021seeing}—the diverse identities, backgrounds, and lived experiences that shape how individuals perceive and respond to cyber threats. For example, in particular, women and gender-diverse communities, and culturally and linguistically diverse (CALD) communities are disproportionately affected by cybersecurity vulnerabilities \cite{cross2022exploring}. In early 2023, phishing scams through various social engineering techniques resulted in over $5\%$ million in losses, representing a diverse age group from CALD communities - women are more susceptible for phishing than men. Australian women, gender-diverse individuals, and culturally and linguistically diverse (CALD) communities are often more susceptible to phishing and other forms of cybercrimes due to factors such as language barriers, limited access to cybersecurity education, and social isolation \cite{leyton2021culturally}. These vulnerabilities can make it easier for attackers to exploit them through deceptive communications, making them even more susceptible to phishing scams and other forms of cyber-attacks \cite{Anti-Scam}.

In addition, these communities face significant barriers to entry and advancement within the cybersecurity workforce, including gender bias, cultural discrimination, and lack of representation in leadership roles \cite{anwar2017gender, namukasa2023understanding, peacock2017gender}. The lack of gender inclusivity, coupled with inadequate attention to other forms of such CALD diversity \cite{sheanoda2024sexuality, leyton2021culturally}, results in a workforce that lacks the broad perspectives needed to address today’s multifaceted cybersecurity challenges.

Despite these complexities, Australia's Higher Education (HE) sector remains anchored to a cybersecurity curriculum that is largely one-dimensional \cite{griffincybersecurity}, overlooking the importance of intersectionality in student engagement and learning \cite{griffincybersecurity}. Current programs tend to emphasise technical proficiency over cultural and social contexts, leaving critical blind spots in how students from diverse backgrounds may respond differently to threats like phishing (social engineering attacks) and other forms of cyber-attacks \cite{griffincybersecurity, farlowcase}. For example, students from marginalized communities may face unique challenges due to language barriers, cultural nuances in communication, or varying levels of access to digital resources \cite{oropeza2010linguistic, thomas2021australian}—factors that are not adequately addressed in the existing curriculums in Australian HE.
Therefore, a pressing need has emerged to revamp the Australian HE cybersecurity curriculum, fostering a more culturally inclusive and comprehensive approach that actively engages students by reflecting their diverse backgrounds, experiences, and perspectives \cite{farlowcase}. By integrating intersectionality into cybersecurity education, we can bridge these skill gaps, cultivating a generation of cybersecurity professionals who are not only technically proficient but also culturally competent to combat against cyber-attacks.

To the best of our knowledge, there is little empirically known about current strategies employed to embed women, gender diverse and CALD perspectives into Australian HE cyber security curriculums,
barriers for doing so, and future work and support needed.  Motivated by this need  we conducted an empirical investigation using in-depth semi-structured interviews with 15 academics from multiple Universities across all states in Australia.  Our study findings contribute to the state-of-the-art understanding of by
(i) providing an evidence-based understanding of the as-is state of embedding women, gender diverse and CALD perspectives into Australian HE cyber security curriculums, (ii) identifying the barriers academics face in embedding women, gender diverse and CALD perspectives into Australian HE cyber security curriculums, (iii)  discussing of the  future work and support needed in this context.  

The rest of the paper is organised as follows. Section~\ref{relatedwork} provides a comprehensive summary of the related work and  Section~\ref{studymethods} gives detailed analysis of the methods employed for data collection and analysis. Section~\ref{studyfindings} presents the findings arising from the qualitative data gathered through the study and the paper ends with a discussion of findings, threat to validity and conclusion in Section~\ref{threats} and Section~\ref{dicussion}.

\section{Related work} \label{relatedwork}
\subsection{Cybersecurity Education in Australian Higher Education}

Cybersecurity has rapidly evolved from a niche technical domain into a pervasive societal concern encompassing human, social, and cultural dimensions \cite{creese2021social}. Despite this shift and the growing recognition of cybersecurity as a multidisciplinary field, higher education curricula remain largely technical in focus and fragmented in scope. Vykopal et. al., \cite{vykopal2025cybersecurity} show that many university cybersecurity programs worldwide fail to sufficiently integrate non-technical dimensions, including human, organisational, and societal factors, as well as broader contextual competencies. While this work highlights a significant lack of interdisciplinary integration, it stops short of examining how intersectional perspectives—such as gender, culture, socio-economic status, and the experiences of marginalised groups—are incorporated into cybersecurity education. This omission points to a critical knowledge gap \cite{vykopal2025cybersecurity}: the absence of systematic integration of intersectionality within curriculum design. Addressing this gap is essential to ensure that future cybersecurity professionals are equipped to engage with diverse user contexts, understand digital inequalities, and respond effectively to the uneven impacts of cyber risks across populations.

Furthermore, the Australian Cyber Security Strategy 2023–2030 highlights a growing gap in cybersecurity capabilities, especially in areas that rely on understanding human vulnerabilities, such as phishing and social engineering attacks \cite{abeysooriya2023discussion}. Despite this evolution, most higher education (HE) cybersecurity curriculums (e.g., NICE (National Initiative for Cybersecurity Education) framework \cite{newhouse2017national}, the Bologna standard \cite{de2018bologna} etc. — broadly adopted and used as a standard guideline for the formulation of cybersecurity degrees) remain technologically deterministic and lack critical engagement with socio-cultural factors that shape digital risks and safety behaviours \cite{abeysooriya2023discussion, mountrouidou2019securing}.

Globally, computing education researchers have highlighted the need for a broader and more inclusive approach to cybersecurity pedagogy \cite{parrish2018global, mountrouidou2019securing}, as reflected in the principles of Culturally Responsive-Sustaining Computer Science Education \cite{center2021culturally} in general. This includes not only emphasizing technical proficiency but also fostering cultural competence, ethical reasoning, and user-centered thinking in cybersecurity practice. Within this context, human-centered cybersecurity education frameworks have emerged \cite{straight2024decentering, straight2024beyond}, recognising that ``users—not systems—are'' often the primary targets of attack \cite{arachchilage2014security}.

Cybersecurity education in Australia is shaped by a layered ecosystem of industry skills frameworks, accreditation standards, and curriculum guidelines across schooling, vocational training, and higher education \cite{griffin2024cybersecurity}. At the professional level, the Skills Framework for the Information Age (SFIA) is the dominant industry standard, defining ICT and cybersecurity roles, skill levels, and responsibilities, and guiding workforce development across government and industry.
Aligned with SFIA, several Australian education-focused frameworks shape curriculum design but remain largely skills- and competency-driven. The Australian Computer Society (ACS) Core Body of Knowledge (CBOK) underpins accreditation of university computing and IT degrees, mapping academic learning to industry expectations and migration pathways \cite{griffin2024sfia}. While CBOK includes ethics and professionalism, these aspects are typically addressed at a high level and often confined to single courses, rather than embedded across Australian HE cybersecurity curricula.

On the other hand, the Australian Digital Capability Framework (ADCF), adapted from the European DigComp framework and adopted nationally from 2025, defines digital skills across five domains spanning foundational literacy to advanced ICT competence \cite{dewr_adcf_2025}. Although the ADCF provides a structured progression of digital capability across Vocational Education and Training (VET) and workforce training, it focuses primarily on individual skill acquisition and does not explicitly address how cybersecurity risks and practices intersect with gender, culture, or language \cite{dewr_adcf_2025}. On the other hand, the ASD Cyber Skills Framework, developed by the Australian Signals Directorate (ASD), aligns directly with SFIA to define cybersecurity roles and proficiency levels for education and workforce development \cite{asd_cyber_skills_framework_2020}. While this framework strengthens career pathways and role readiness, its emphasis remains on technical capability and professional progression, offering limited guidance on inclusive or culturally responsive education \cite{dewr_adcf_2025, asd_cyber_skills_framework_2020}. At the pre-tertiary level, the Australian Curriculum: Digital Technologies (F–10) and senior secondary computer science syllabuses emphasise computational thinking, problem-solving, and practical skills, adopting a competency-based orientation similar to SFIA \cite{acara_digital_technologies_f10}. However, these curricula largely model a generic learner and user, with minimal engagement with social, cultural, or intersectional dimensions of digital technologies and cybersecurity.

Collectively, these frameworks demonstrate strong alignment between education and industry needs but share a common limitation: inclusivity and intersectionality are implicit values rather than explicit curriculum design principles, resulting in uneven and superficial integration within cybersecurity education \cite{griffin2024sfia, dewr_adcf_2025, vykopal2025cybersecurity, accenture2026cyberworkforce}.

\subsection{Intersectionality and Digital Vulnerability}
Coined by Kimberlé Crenshaw, intersectionality explains how overlapping systems of oppression shape individuals' lived experiences \cite{Crenshaw1989}. In the context of cybersecurity education, intersectionality becomes critical when examining how identity characteristics—such as gender, ethnicity, socio-economic status, and language—compound digital risk exposure and influence users’ responses to cyber threats \cite{MittaDupuis2024, vykopal2025cybersecurity}. 

Evidence from Australian contexts shows that women, gender-diverse individuals, and culturally and linguistically diverse (CALD) populations are disproportionately affected by phishing related scams, romance fraud, and technology-facilitated violence and abuses \cite{eSafety2019, HenryFlynn2018, ACCC2024}. Attackers often exploit cultural norms/standards, communication styles, values, or expectations to enhance the effectiveness of phishing and social engineering cyber crimes \cite{desolda2021human}. For example, in high power distance cultures—where hierarchy and authority are potential—attackers may impersonate senior executives to pressure employees into urgent actions, such as authorizing payments or disclosing credentials \cite{gottschalkpolicing}. This tactic, known as business email compromise (BEC) \cite{al2023business, bonhard2025review} in the organisational settings, thrives on culturally reinforced obedience to authority figures. For instance, cybercriminals may impersonate a Chief Executive Officer within an Asian multinational organisation and exploit culturally embedded hierarchical communication norms. By adopting formal, authoritative language and leveraging perceived positional power, attackers can coerce employees into complying with fraudulent requests—such as initiating unauthorised fund transfers—often before adequate verification occurs \cite{abawajy2014user}. Similarly, in collectivist cultural contexts, attackers frequently appeal to notions of group loyalty, organisational harmony, or community obligation to manipulate targets emotionally. Such culturally tailored social engineering attack strategies increase perceived trust and reduce scepticism, thereby heightening individuals’ susceptibility to deception \cite{workman2008wisecrackers}. Collectively, these examples underscore the critical need for culturally and linguistically aware (i.e., CALD) cybersecurity education and training to strengthen resilience against sophisticated social engineering attacks. Such threats are often deliberately tailored to exploit culturally embedded norms, values, and communication practices, making CALD communities disproportionately vulnerable. Addressing these risks requires cybersecurity interventions that move beyond generic awareness approaches and instead account for cultural context, language, and social dynamics within diverse populations. Furthermore, these communities also face systemic exclusion from cybersecurity education and workforce participation due to structural inequities, such as limited access to technical training, lack of culturally safe curriculums, and implicit biases in pedagogy and hiring practices \cite{eSafety2019, APSC2024CALDstrategy}. Yet, intersectionality is rarely operationalized in HE Australian cybersecurity education. Curriculums typically overlook the nuanced ways in which marginalised learners experience, understand, and respond to digital risks, leading to educational practices that are exclusionary, disengaging, and ineffective for diverse learners.

\subsection{Current Limitations in Australian Cybersecurity Curriculums}
Griffin and Johnson \cite{griffin2024sfia} critically examined how Australian HE cybersecurity programs align with the Skills Framework for the Information Age (SFIA), revealing a skewed emphasis toward technical knowledge at the expense of socio-ethical, psychological, and user-centered skills
. This narrow framing fails to engage students who bring valuable lived experience with digital marginalization or who require culturally responsive teaching approaches to thrive (i.e., in our case women, gender-diverse and CALD community). Moreover, cybersecurity curricula and pedagogical practices frequently adopt a deficit‑oriented framing when engaging with diverse student cohorts. In this approach, learners from non‑normative identities—such as women, culturally and linguistically diverse (CALD) backgrounds, and gender‑diverse communities—are positioned as inherently “at risk” or lacking, rather than being recognised for the distinctive skills, perspectives, and strengths they bring \cite{valencia2012evolution, mountrouidou2019securing}. Such framings risk reinforcing structural inequities by marginalising diverse forms of knowledge and limiting meaningful participation in cybersecurity education and the broader workforce. Such perspectives can reduce motivation, lower self-efficacy, and limit career aspirations among under-represented students \cite{ladson1995toward} in cybersecurity within a country. As a result, there is growing consensus that HE institutions must transition from generic technical training to inclusive, intersectionality-informed curriculums that reflect the realities of digital life for all students—not just a normative, predominantly male, English-speaking, and Western cohort \cite{mountrouidou2019securing}.

\subsection{Toward Intersectionality-Informed Cybersecurity Pedagogy}
Addressing these disparities requires a pedagogical shift toward inclusive, intersectionality-informed teaching practices \cite{mountrouidou2019securing}. These may include: Embedding culturally responsive teaching materials that reflect students' lived experiences. Incorporating collaborative learning and storytelling approaches that allow marginalized voices to shape content. Designing assessments that consider diverse linguistic and cultural modes of knowledge representation. Fostering supportive peer networks and mentorship opportunities for under-represented groups. Most importantly, such curriculums reforms must be informed by evidence based research, and the voices of those most affected by educational exclusion. Therefore, co-designing with students from marginalized communities—especially women, CALD, and gender-diverse individuals—ensures that reforms are grounded in real-world needs, not institutional or individual assumptions.

\subsection{Gaps in the Literature and Research Contribution}

Australia’s demographic profile has shifted markedly in recent years, with Census data showing that people born in India now constitute the largest overseas‑born population, alongside substantial growth from countries such as China, Nepal, the Philippines, Vietnam, and Sri Lanka \cite{ABS2025CountryBirth}. This increasing cultural and linguistic diversity is especially visible within Australian Higher Education and the cybersecurity workforce pipeline. However, Australian HE cybersecurity curricula continue to be designed around a largely homogeneous, English‑dominant learner model, with limited consideration of how migration background, language, and cultural norms shape digital practices and cyber risk \cite{griffin2024cybersecurity, griffincybersecurity, griffin2024sfia, dewr_adcf_2025, asd_cyber_skills_framework_2020, acara_digital_technologies_f10}. As Australia’s population and talent base become increasingly multicultural \cite{ABS2025CountryBirth}, the absence of culturally responsive and intersectionality‑informed cybersecurity education represents a growing structural misalignment between who is being educated and how cybersecurity is taught.

On the other hand, despite increasing attention to diversity in computing education \cite{hayes2026principle}, very few studies have operationalized intersectionality in cybersecurity curriculum design \cite{mountrouidou2019securing}. Most research focuses either on diversity in the workforce or on digital literacy interventions in community settings—not on transforming university HE curriculums to centre intersectional experiences within technical disciplines \cite{namukasa2023understanding}.

Despite the existence of multiple, well-established skills and education frameworks in Australia, there is little empirical research examining how intersectional perspectives are—or are not—embedded within cybersecurity curricula aligned to these frameworks \cite{griffin2024cybersecurity, griffincybersecurity, griffin2024sfia, dewr_adcf_2025, asd_cyber_skills_framework_2020, acara_digital_technologies_f10}. Existing studies typically focus on workforce shortages, diversity in employment, or digital inclusion initiatives outside Australian Higher Education, rather than interrogating curriculum design practices within accredited university programs. Therefore, this study addresses this gap by providing the first multi-university empirical investigation of how Australian HE cybersecurity curricula currently engage with women, gender-diverse, and CALD perspectives, and by identifying structural, institutional, and pedagogical barriers that limit meaningful inclusion. By situating these findings within the broader ecosystem of SFIA-aligned frameworks (ACS CBOK, ADCF, ASD Cyber Skills Framework, and national curricula \cite{griffin2024cybersecurity, griffincybersecurity, griffin2024sfia, dewr_adcf_2025, asd_cyber_skills_framework_2020, acara_digital_technologies_f10}), this research offers actionable insights for educators, accreditation bodies, policymakers, and industry seeking to strengthen both workforce relevance and educational inclusivity.

This proposed research fills this gap by: i) conducting a multi-university analysis of current cybersecurity offerings across Australian HE institutions; ii) applying co-design methodology with students from intersecting marginalized identities; iii) developing practical, adaptable guidelines for embedding intersectionality in cybersecurity education across the Australian HE institutions. This contribution is especially timely given the broader societal push for equity in STEM education, the urgency of human-centred cybersecurity threats, and the increasing policy focus on inclusive digital capabilities in Australia.

\section{Methods} \label{studymethods}

We used semi-structured interviews to gather data from academics who have experience in teaching and or coordinating cyber security courses at Australian universities. These interviews explored their current strategies for embedding women, gender-diverse and CALD  perspectives into the Australian HE cyber security curriculum, barriers faced, and the reasons for either integrating or not integrating intersectional perspectives into the Australian HE cybersecurity curriculum, and future direction and support needed. All participants received a 40 AUD gift card as a token of appreciation for their contributions.   Ethical approval was obtained from the Human Research Ethics Committee of [removed to maintain anonymity]  University prior to commencing data collection.

\subsection{Interview guide preparation} \label{interviewguide}

The interview guide had three main sections: i) current strategies for inclusivity; ii) barriers to embedding women, gender-diverse, and CALD perspectives into Australian HE  cyber security curriculums; iii) future work and support needed.  All authors contributed to the development of the interview guide. A  pilot was conducted to identify potential issues with question clarity and phrasing.  The pilot interview helped us to understand two issues in the way we originally planned to ask the interview questions: i) not all participants may be fully aware of the key terms used in the interview guide, and ii) participants may be hesitant to disclose their personal challenges but more willing to articulate them as general issues. As a result, we decided to include: i)  an introductory slide set about the project and the key terms used during the project (e.g., gender-diverse, CALD, intersectionality); ii) explanation at the start that participants are not required to base the responses solely on personal experiences; they may also draw upon current/past observations or what they have heard etc. We also changed the wording of certain interview questions to reflect these changes.    The anonymised introductory slides and the interview questions are available here\footnote [1]{\href{https://osf.io/gpyek/?view_only=b5c54b0ae13248ad833f20901226f589} {Interview material is available here.}}.

\subsection{Recruitment and data collection procedure}
We recruited 15 academics who had experience in teaching and/or coordinating cybersecurity courses in the Australian higher education sector.  These academics included three Professors, four Associate Professors, three Senior lecturers and five lecturers representing all states of Australia. The demographics of the participants are given in Table~1. Participants were recruited via invitation emails sent to potential study candidates identified through existing contacts or web searches. Most of these participants had extensive experience teaching cybersecurity across various universities within the Australian higher education sector.

The interviews were conducted using Microsoft Teams and video recorded for analysis and transcription purposes. The interview lasted approximately 45-60 minutes. Two researchers were present in each interview. One researcher primarily led the discussion, while the other took detailed notes. At the start of each interview, we provided participants with an overview of the project, the aims of the interview study, explanations of the expectations from them, and an explanation of how the data would be analysed and used, using the introductory slides deck. We also provided them with the opportunity to ask the researchers any questions they had before the interview began.  After that, participants were asked to provide answers to a set of basic demographic questions. A researcher shared their screen with a Google form for this purpose and filled in the data as the participants explained their answers to the given questions (e.g., years of experience teaching cyber security at the undergraduate level). When the participants were satisfied with their answers to the basic demographic information requested, the researcher submitted the form on their behalf.   Next, we moved on to the interview questions. 
The interview was guided based on the interview guide explained in Section~\ref{interviewguide}. At the end of the interview, participants were provided the opportunity to express any additional points they wished to make beyond the interview questions. Finally, the researchers thanked the participants for their time and mentioned that a gift card would be sent to them as a token of appreciation after the interview.

\subsection{Data analysis procedure}
\begin{figure*}
\centering
            \includegraphics[width=0.7\linewidth]{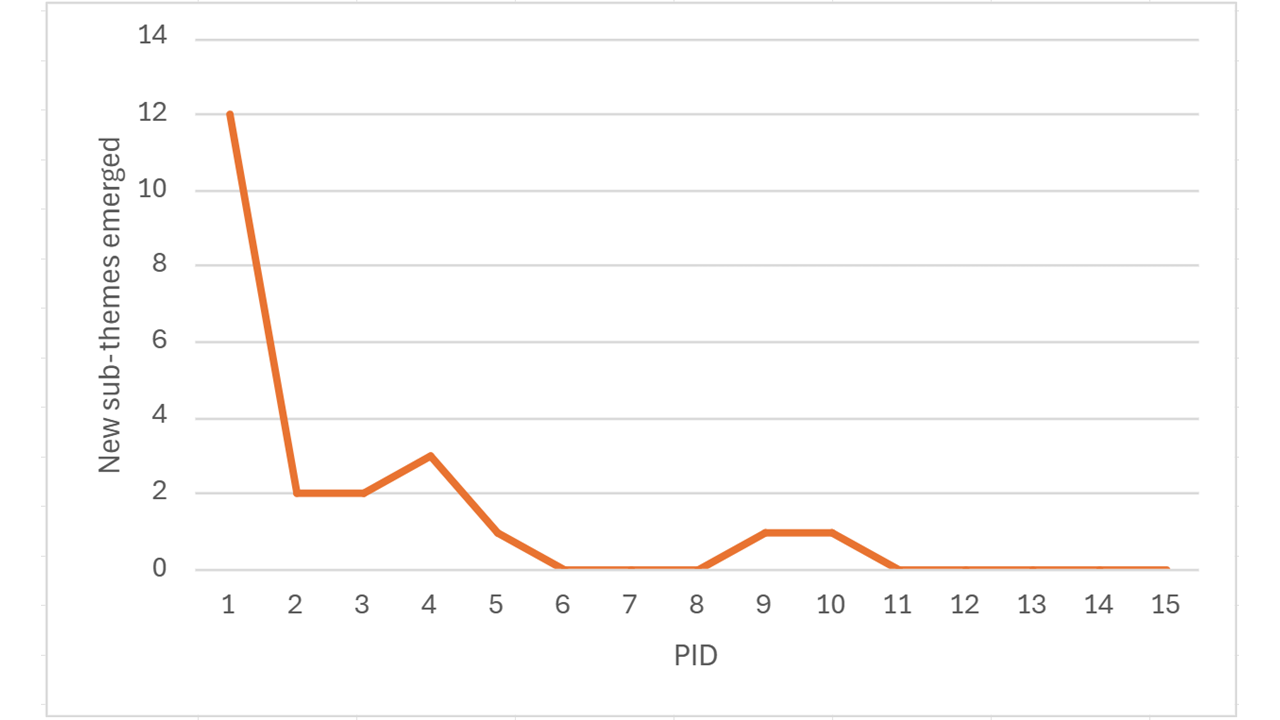}
        \caption{Theoretical saturation achieved during data analysis. The last five interviews provided more examples for the emerged findings but no new themes  emerged. }
        \label{saturation}
\end{figure*}

We analysed the qualitative data obtained from interviews using thematic analysis based on the guidelines proposed previously \cite{braun2006using}. Two authors, who had more than 10 years of experience with qualitative data analysis, worked closely together during the analysis. The process
involved: i) familiarising with the data; ii) generating initial codes; iii) searching for themes; iv) reviewing themes; and
v) defining and naming themes. We initially performed a pilot data analysis using data from three participants and discussed the findings before proceeding with the rest of the analysis. After the pilot, one researcher primarily led the analysis and held frequent meetings with the other researchers to review progress. Any disagreements were resolved through discussions during these meetings. We achieved theoretical saturation (i.e., a state where little fresh information emerges from subsequent interviews) because all the main themes and sub-themes had been uncovered across the 15 participants (see Figure~\ref{saturation}). For example, the last five interviews provided more examples for the emerged findings but no new themes or insights emerged.

\begin{table*} 
  \caption{Participants demographics}
  \label{tab:participantsdemographics}
  \centering 
  \small
  \begin{tabular} {p{.02\textwidth }  p{.15\textwidth} p{.15\textwidth} p{.11\textwidth}  p{.11\textwidth} p{.11\textwidth} p{.11\textwidth} }
    \toprule
    PID &  State &Current  \newline{title}  & Yrs of  \newline{Exp} \newline{teaching cybersecurity  at the}\newline{postgraduate level} & Yrs of  \newline{Exp} \newline{teaching cybersecurity at the}\newline{undergraduate level} & Yrs of  \newline{Exp} 
    \newline{coordinating }\newline{postgraduate cybersecurity subjects}
     &  Yrs of  \newline{Exp} \newline{coordinating }\newline{undergraduate cybersecurity  subjects}  \\
    \midrule
P1 & New South Wales & Associate Professor & 9 & 9 &  9  &  9\\ 
P2 & Tasmania & Lecturer  & 5 & 16 &  1  &  1\\ 
P3 & New South Wales & Senior Lecturer  & 6 & 10 &  4  &  4\\ 
P4 & New South Wales & Professor   & 6 & 6 &  6  &  6\\ 
P5 & South Australia & Professor   & 20 & 25 &  15 & 18  \\    P6 & South Australia & Lecturer   &5 & 0 &  5  & 0  \\   
P7 & Queensland  &Associate Professor   & 4  & 4 &  4 & 4  \\ 
P8 & South Australia   &Associate Professor   & 6  & 0 &  6 & 0  \\
P9 & Victoria  &  Professor   & 20  & 20 &  20 & 20  \\
P10 & Western Australia & Senior Lecturer & 6 & 6 & 4 & 5 \\
P11 & New South Wales &  Lecturer & 5& 1 & 5 & 1 \\
P12 &  Tasmania &   Lecturer  & 5 &  4 & 5  & 4  \\

P13 & Victoria & Associate professor  &  &  &  &  \\
P14 & New South Wales &  Lecturer & 5 &  3 &  5 &  3\\
P15 & Tasmania & Lecturer  & 5 & 5   & 0   & 5  \\

    \bottomrule
  \end{tabular} \label{demgraphics}
  
  \smallskip
  \end{table*}

\section{Findings} \label{studyfindings}
\begin{figure*}
\centering
            \includegraphics[width=1\linewidth]{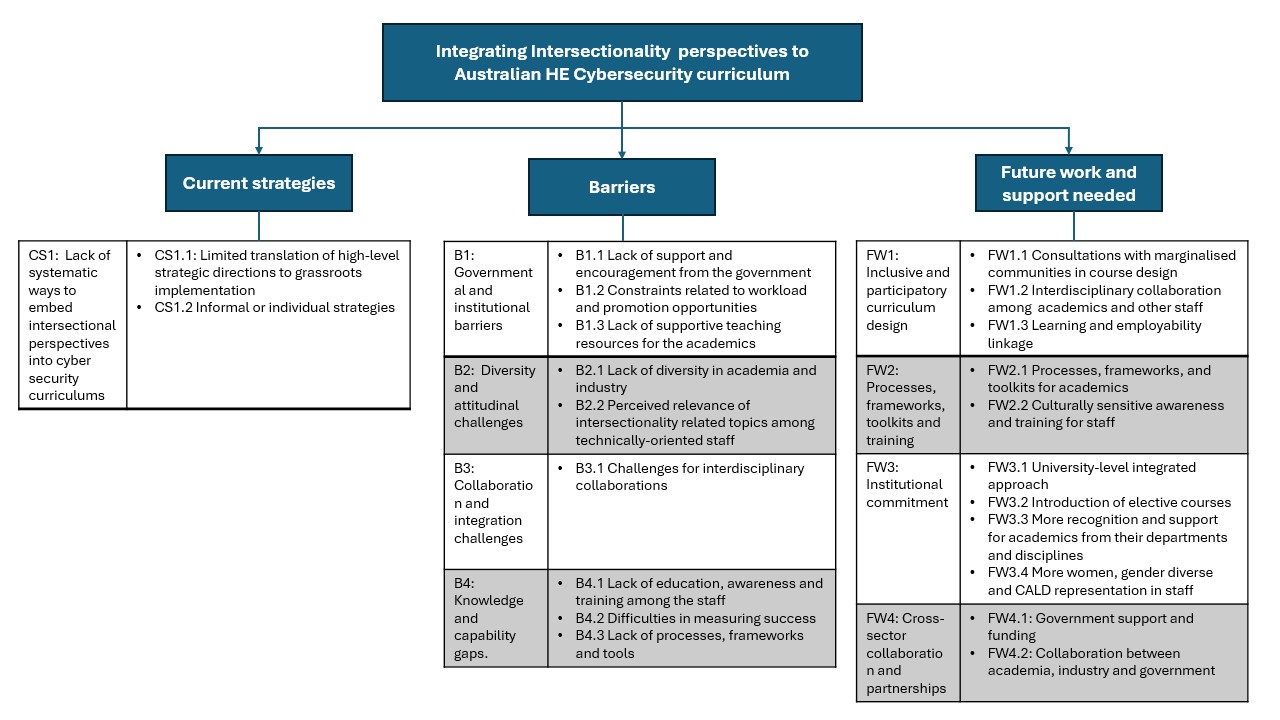}
        \caption{Overview of the study findings}
        \label{findings}
\end{figure*}

The semi-structured interviews conducted with academics who had experience in teaching and/or coordinating cybersecurity courses in the Australian higher education sector provided detailed insights into current strategies for embedding women, gender-diverse and CALD perspectives into cyber security curriculums, barriers for doing so and the future directions and support needed, which we explain below. An overview of study findings is presented in Figure~\ref{findings}.


\subsection{Current strategies}

In this section, we present participants' explanations about their current strategies to embed intersectional perspectives, specifically women, gender-diverse and CALD perspectives, into cyber security Australian HE cyber security curriculums. 

\subsubsection{CS1: Lack of systematic ways to embed intersectional perspectives into cyber security curriculums}

\paragraph{CS1.1: Limited translation of high-level strategic directions to grassroots implementation}
All participants unanimously agreed that, currently, there are limited or no systematic ways to embed women, gender-diverse and CALD perspectives into cyber security curriculums. Although university strategies often emphasise the importance of inclusivity and diversity, participants noted that there is limited support for translating these commitments into grassroots implementation, particularly in curriculum design and delivery  \blockquote[P1]{Most of the universities where I worked, and I am aware that they all have a policy for diversity and inclusion. But it is very unlikely you will see something exclusive for cybersecurity.  You will see that in the industry I work in, I work very closely with the industry and the Community, and you'll see there is a big demand. There is a gap in women's representation in the cyber and cloud communities.  However, we don't see many policies in place}.

In terms of intersectionality, beyond the inclusion of women, gender-diverse, and CALD perspectives, some participants noted that there have been encouraging advancements in integrating Indigenous perspectives into the cybersecurity curriculum. On the other hand, some participants acknowledged the presence of active members who are experts in Indigenisation but noted that these experts have not yet been able to embed Indigenous perspectives into cybersecurity programs or courses.

\paragraph{CS1.2: Informal or individual strategies}
While there are currently no systematic or institution-wide approaches in place for embedding women, gender-diverse, and CALD perspectives into cyber security curriculums, some participants described informal or individual strategies they have adopted to address these issues to some extent in their courses or those taught by people they know. For example, some explained that general discussion about diversity and inclusion is carried out as a part of the ethics related course/topics within the cyber security curriculams.  \blockquote[P9]{In regards to sort of management and society issues, so again, you know you may have a situation where gender is discussed, but it's like it'd be like one week in an ethics course, you know, and that's the extent.} 
Another participant explained that she would do an early assessment in a course she designed, which is considered a diagnostic test to determine who needs support in language. Based on the results of this early assessment, their team would then conduct learning circles to ensure that students lacking language skills are well supported and brought up to a standard that allows them to follow the course.  \blockquote[P5]{We did the first assignment really early, and it was a diagnostic test. So with the diagnostic test, I could actually pick out people whether they. Whether they were older males with no academic experience or whether they were from Vietnam with hardly any English. Then my friend would actually put in the learning circles to bring them to the stage where we believe that they could. Deal with the curriculum as it was. So, generally speaking, we were dealing with language.}


\subsection{Barriers} \label{barriers}
In this section, we describe what participants explained when they were asked to explain barriers for embedding women, gender-diverse and CALD perspectives into cyber security curriculams. 

\subsubsection{B1: Governmental and institutional barriers}

\paragraph{B1.1 Lack of support and encouragement from the government} \label{CH:lackofgovernmentsupport}
Our participants emphasised that the government should increase funding to universities to better support women, gender-diverse, and CALD communities. Currently, there is a noticeable lack of targeted initiatives in these areas, highlighting the need for clearer support structures. 
One participant even highlighted that there is no dedicated ministry representing these minority groups. Due to financial pressure, universities are often compelled to restrict their capacity to provide comprehensive support and promote inclusive environments.   \blockquote [P7]{There is financial pressure in our universities and limitations in providing the funding resources to cover this area. At least for my organisation, I had a discussion yesterday about some updates to the course which is needed. However, we still need to wait until the end of this year, as any course revision, even if agreed upon by the discipline, requires support from the school. But the school doesn't have sufficient financial support}. Additionally, when the governments change, policies and funding priorities often shift, which can disrupt ongoing initiatives and create uncertainty for universities striving to support diverse communities effectively \blockquote [P4]{Is it important that Universities really want this to happen, even if when things when times get tough, I don't want to be this the first one to drop off when there is a workplace change. I've seen this, which is not good}. Participants highlighted that by addressing these issues, Australia can ensure its higher education institutions become more equitable and effective.

\paragraph{B1.2: Constraints related to workload and promotion opportunities}

 Participants noted that their workloads frequently leave little to no time to explore the importance or relevance of intersectional perspectives in their courses and/or to integrate these perspectives into their teaching.  As a result, the demanding nature of academic responsibilities often hinders engagement in diversity and inclusion initiatives. It restricts participation in activities beyond core teaching and research duties that are prioritised through the workload model.  Furthermore, when academics perceive that these initiatives are not properly recognised in their promotion pathways, their motivation to engage in them diminishes further \blockquote[P13]{When they [academic staff]  apply for a promotion, then that [collaborative work done in designing courses with intersectional perspectives] should be highly valued.  It is easy to say that - We encourage cross-university, cross-college, cross-departmental collaborations, but at the same time, the value should be heard}.

\paragraph{B1.3: Lack of supportive teaching resources for the academics}

The lack of teaching resources is one of the significant barriers for them to integrate women, gender-diverse and CALD perspectives into the cyber security courses they teach. In fact, our participants identify two specific gaps. Firstly, there is a lack of research that informs curriculum design with robust scientific underpinnings in the areas of diversity and inclusion, making it challenging to develop evidence-based educational content. Secondly, there is a shortage of relevant use cases and personas that can be utilised in classrooms as examples. This limits the ability to engage students with practical, real-world scenarios that reflect diverse perspectives \blockquote[P6]{The challenge is we do not have access to some sort of pool of real-world use cases or case studies so that you can bring them into either ethical discussions or sort of human-centric courses}.

\subsubsection{B2: Diversity and attitudinal challenges}

\paragraph{B2.1: Lack of diversity in academia and industry}
Cyber security field is predominantly dominated by males, which creates a lack of diverse perspectives and experiences within the field. In particular, having more women and gender-diverse and CALD representation in academia and industry potentially creates more opportunities for cyber security solutions and approaches that a more diverse workforce could bring. This diversity could also influence the development of curriculums, ensuring that educational content reflects a broader range of perspectives and prepares students to address complex, real-world challenges. Participants explained that cyber security academics, who are often males, frequently use case studies and examples biased towards males and mono-cultures in their lecture material and assignments \blockquote{Often when we write assignments and exams, we say, imagine an organisation with, you know, so, so many people working on. Often, the examples that we give are very male and very mono-cultural}.  On the other hand, from the students' perspective, having fewer women, gender-diverse, and CALD role models could limit their inspiration to pursue courses and careers in the cyber security field and to excel in them. 

\paragraph{B2.2: Perceived relevance of intersectionality related topics among technically-oriented staff}

Cyber security is inherently considered a technical subject, and most academics teaching cyber security courses in the HE sector possess the most experience with technical skills such as coding, network security, and encryption methods. They also focus on teaching these skills to students. \blockquote [P4]{So who's going to teach it? Most of the teachers are actually the  hardcore cybersecurity people.}

Furthermore, one participant explained that for master's level courses, students come from diverse backgrounds; hence, often providing technical skills and knowledge is seen as more critical by both educators and students.  Technical skills are crucial for a graduate completing an undergraduate or master's degree related to cyber security; however, to identify and develop effective solutions for today's multifaceted cyber security challenges, the students should be equipped with a much broader skill set. This includes not only technical expertise but also critical thinking, ethical reasoning, and an understanding of the socio-cultural dimensions of cyber security. Unfortunately, most technically oriented academic staff seem to overlook this fact and are reluctant to incorporate intersectional perspectives into the courses they teach.

\subsubsection{B3: Collaboration and integration challenges}
\paragraph{B3.1 Challenges for interdisciplinary collaborations}

Our participants emphasised the need for interdisciplinary collaborations for embedding women, gender-diverse and CALD perspectives into the cyber security curriculum. For example, cyber security academics working hand in hand with faculty members from the School of Psychology were identified as a possible step towards identifying effective ways to embed intersectional perspectives into the learning materials and assessments.  Although interdisciplinary approaches offer significant benefits, participants explained several challenges in their implementation, which we describe below. 

Participants explained that the different mindsets across diverse disciplines present challenges in both teaching and research. In terms of research, different disciplines favour different publication venues. The divergence in publishing priorities can make it difficult for researchers from various fields to find common ground and effectively share their interdisciplinary work with broader audiences \blockquote[P13]{We encourage people to collaborate, but at the end of the day we are publishing in two different venues that have different recognitions... That creates a huge barrier for people to collaborate hand in hand with the other people}.

Similar issues can arise in teaching. The focus of two or more disciplines on teaching matters could vary significantly, creating unnecessary tension. This divergence may lead to conflicts over curriculum priorities, teaching methodologies, and assessment criteria. As a result, cross-disciplinary collaboration needs significant time and effort. However, participants fear that they may receive limited recognition from top management for such interdisciplinary work. 

\subsubsection{B4: Knowledge and capability gaps} \label{ch:knowlegeAndCapability}

\paragraph{B4.1: Lack of education, awareness and training among the staff}

Our participants described various limitations in awareness, education and training among the staff, which hinders their capacities to embed women, gender diverse and CALD perspectives into cyber security curriculams. Some participants even acknowledged that they have not thoroughly considered the importance of integrating intersectional perspectives into their teaching practices. This lack of reflection can stem from being focused on traditional academic routines or facing the pressures of immediate teaching and research demands, leaving little room for exploring novel avenues that enrich students' overall learning experiences. 

On the other hand, even the academics who are aware of the importance of embedding intersectional perspectives into cyber security curriculum, lack of education and training on how to embed these perspectives into their courses and to make the integration a relevant way for students \blockquote[P11]{familiarity with the area would also be something that would hold them [academics] back. I think there might be a lack of education there. There is a general understanding that diverse perspectives are important. But there is a lack of knowledge of exactly  what  those different perspectives  and tools are and how to integrate them into their teaching}.

Furthermore, some participants explained they or other academic staff they know lack the confidence to discuss sensitive topics in the classroom. They are concerned that the examples or vocabulary they use could offend a particular student or even a group of students.  

\paragraph{B4.2: Difficulties in measuring success}
Measuring success in incorporating intersectional perspectives into curriculams requires more than just tracking the increase in enrolment from diverse student groups. A participant explained that it should involve evaluating how well these perspectives are integrated into learning outcomes and assessing their impact on students' learning experience. However, until now, there are no clear ways of measuring success in tangible and fair ways. \blockquote[P4]{The quality is often hard to measure. Often, it does not appear well in Excel sheets. Yes, you can measure - how many females? These are things you can measure. However, some things cannot be measured. What is your goal, and what is your target? What are the measurable outcomes of achieving much better inclusivity and diversity?}

Even a participant explained that their University is putting a lot of effort into embedding indigenous-specific perspectives, issues, and solutions into the courses they teach; however, they are struggling to identify the most effective ways to measure the success of these initiatives. The participant further explained that the University can show statistics indicating that enrolments from Indigenous students have significantly increased after their initiatives were launched, but this measure alone is not sufficient to identify how effectively these students are engaging with and benefiting from the curriculum or how students, in general, are equipped to tackle real-world cybersecurity challenges upon graduation after the initiatives are introduced.

\paragraph{B4.3: Lack of processes, frameworks and tools}

Our participants explained various issues related to processes, frameworks and tools restricting the ability or demotivating the Universities and academics to focus beyond technical subjects within the cybersecurity curriculums. 

One of the significant factors is how the cyber security curriculams are accredited at the moment. Australia's cyber security curriculums is often accredited based on the Australian Computer Society (ACS), as it provides pathways for international students to immigrate to Australia.   Currently, based on ACS accreditation, the focus is on teaching several core courses that encompass the body of knowledge. These courses are largely technical, with a limited focus on social and human aspects, such as intersectionality.  As a result, most cybersecurity curriculums address inclusivity-related topics only at a surface level, often within an ethics course  \blockquote[P9]{In the Australian context, all IT and information systems courses are credited by the Australian Computer Society.  So the reason universities do that is, again, it's a professional body.  Additionally, it provides a pathway for international students to migrate. So it's very attractive. The Australian Computer Society has a body of knowledge that should be taught in every IT course.  And again, the focus is very much on technology with a very small part focus on ethical,  social and business issues}.


Participants pointed out other countries that have processes and frameworks that focus on gender and inclusivity as a part of the cybersecurity curriculums. For example, the United Kingdom (UK) has developed a cybersecurity body of knowledge (i.e., CyBOK) that universities are expected to teach, along with cyber apprenticeship programs. These initiatives explicitly focus on intersectional perspectives, such as gender, ensuring they are integrated into the curriculum. 
Often Australia tend to follow processes and frameworks that the UK develops. However, a participant highlighted issues associated with adopting a model from another country, such as the UK, and applying it directly to Australia. They noted that each country has its unique characteristics and challenges, which must be taken into account. Australia's distinct cultural and socio-economic landscape can significantly influence the effectiveness of such models. For example, the participant emphasised the importance of tailoring approaches to meet Australia's specific needs and circumstances, rather than assuming that processes, frameworks, and tools successful in other contexts will be universally applicable. 

\subsection{Future directions and support needed}
In this section, we describe participants' perspectives on future work and support needed for embedding women, gender-diverse and CALD perspectives into cyber security curriculums in the Australian HE sector. 

\subsubsection{FW1: Inclusive and participatory curriculum design}

\paragraph{FW1.1: Consultations with marginalised communities in course design}

Participants explained the need for consultations with marginalised communities when designing the Australian HE cyber security curriculum  \cite{eSafety2019, APSC2024CALDstrategy}. Such consultations are crucial for ensuring the curriculum reflects diverse perspectives and addresses the specific needs and challenges faced by these groups. By incorporating insights from marginalised communities, educators can create more inclusive and relevant course content. For example, by consulting with marginalised communities, academics can identify real-world case studies and examples that they can use in class for teaching and assessment.  Having access to such rich real-world case studies and examples will potentially address the challenges related to the lack of supportive teaching resources for the academics described in Section~\ref{CH:lackofgovernmentsupport}. Furthermore, this approach not only enhances the educational experience but also prepares students to consider social and ethical dimensions in their future cybersecurity careers \blockquote [P8]{I definitely think there is a huge opportunity to involve students from some of these specific groups. Co-design curriculums that are relevant to them. So, it's not just curriculum content, but also the delivery that's relevant for them. The reason why that's important is that sometimes it's easy to sit back, especially as academics and researchers, and think, These are your issues. These are your challenges and this is how you fix your challenges. Co-design is so powerful because it's empowering. It's participatory.
It allows people to own whatever solutions they come up with. So for me, yeah, co-design actually will go a very long way in addressing this problem}.

\paragraph{FW1.2: Interdisciplinary collaboration among  academics and other staff}

Collaboration among interdisciplinary academics and other staff in course design is crucial for fostering innovation and enhancing educational outcomes.

In terms of teaching, our participants emphasised the importance of academics teaching cybersecurity courses, collaborating with academics from other disciplines when designing curriculams and course materials. This interdisciplinary approach ensures that courses are comprehensive, reflecting the complexities of modern cybersecurity challenges. By integrating diverse expertise, we can develop cyber security programs that are not only technically robust but also consider intersectional perspectives, preparing students for real-world applications. Such collaborations should be valued, recognised  and promoted from the University management.  Furthermore, one participant suggested the possibility of having a separate team that specialises in intersectional perspectives that could work hand in hand with academics to ensure that diverse perspectives are integrated into the curriculum, leading to more comprehensive and engaging learning experiences \blockquote [P13]{It is something like when you're developing a course, it would be quite nice to have a dedicated team who are working hand in hand with academic. That needs to be properly formulated with awareness of all the schools and heads of schools}.

In terms of research, participants highlighted that more scientific evidence is needed to explain how intersectional perspectives can be effectively integrated into curriculums. They highlighted the need for studies that demonstrate how existing educational frameworks can be contextualised to accommodate diverse viewpoints. This research is essential for developing an inclusive educational approach that addresses the varied needs of students and prepares them for complex environments.

\paragraph {FW1.3: Learning and employability linkage}

Our participants highlighted that students might resist learning cyber security courses embedded with intersectional perspectives if they are uncertain about the relevance of the teaching content to the cybersecurity challenges they could encounter in their careers after graduation. The participants believe that as long as academics can demonstrate the necessity and relevance of looking beyond technical aspects and into intersectional perspectives for solving complex real-world cybersecurity challenges, students will be more open and motivated to engage with the material \blockquote [P10]{I look at the utmost benefit of the students and then I try to reverse engineer my teaching delivery and whatever I teach in the class. So it's my best interest to incorporate all this in my teaching so that my students can become successful. Our job is to make sure the students become global leaders and global experts not just for a certain geographic location so in reality}.

To enhance the professional relevance of educational content, it is essential that all teaching materials, including diverse perspectives, are explicitly connected to course learning outcomes, specific job roles within enterprise security, and broader industry expectations. By incorporating gender and cultural dimensions into practical, job-focused scenarios, these abstract concepts could become more concrete and accessible for students. This method not only deepens understanding but also prepares students to apply these insights effectively in their future careers. As a result, learners are better equipped to navigate and contribute to an increasingly inclusive and dynamic workforce, aligning their skills with the evolving demands of the industry and broader context in which cybersecurity operates.

\subsubsection{FW2: Processes, frameworks,  toolkits and training}

\paragraph{FW2.1: Processes, frameworks, and toolkits for academics}

Often, governments and universities establish strategies aimed at promoting inclusivity and intersectionality. These strategies are in place to promote educational environments that embrace diversity in all its forms, ensuring that curriculums and University practices reflect a variety of perspectives and experiences. However, without proper processes, frameworks, and toolkits, academics struggle to implement those high-level strategies. 

Our participants pointed out that evidence-based research is needed to determine how to effectively integrate intersectional perspectives into curriculums and how existing frameworks can be contextualised. However, as previously pointed out (see  Section~\ref{ch:knowlegeAndCapability}), adapting frameworks from other countries to Australia may not be straightforward. This complexity arises from Australia's unique cultural and social contexts, which necessitate tailored approaches to ensure these frameworks are relevant and effective in addressing local intersectional challenges. 

A participant also emphasised the importance of engaging in discussions with accreditation bodies, such as the Australian Computer Society, to explore the possibility of updating the body of knowledge and curriculum model to ensure better representation of gender issues and other forms of inclusivity.   \blockquote [P9]{If you go back to that fundamental issue that Australia has with our courses about curriculum design and accreditation, it would actually be important to have a conversation with the Australian Computer Society.  There is an existing body of knowledge that every course is supposed to teach.  It is whether that body of knowledge and that curriculum model.  Can it be updated to reflect gender issues and inclusivity as part of the curriculum of every course?}. As explained in  Section~\ref{ch:knowlegeAndCapability}, if the body of knowledge has an emphasis on intersectional perspectives, there is more motivation for the Universities to actively start looking into ways of embedding these concepts into their curriculums.

Apart from the aforementioned processes and frameworks, academics also need access to targeted resources and examples. These resources should include, but are not limited to, a comprehensive pool of case studies, cybersecurity challenges, and potential solutions \blockquote[P6]{I think it would be great to have a pool of examples. What can we pull or bring into the ethical discussions or human-centric topic courses?}. Such resources would enable educators to illustrate practical applications of intersectional perspectives and equip students with the skills necessary to address complex, real-world cybersecurity issues.

\paragraph{FW2.2 Culturally sensitive awareness and training for staff}

One of the major challenges academics are currently facing, as described in Section~\ref{ch:knowlegeAndCapability}, is the lack of awareness and training. Many academics are not fully aware of the cybersecurity challenges experienced by women, gender-diverse, and CALD  communities. 
Even those who recognise these issues often lack the expertise to integrate these perspectives into their courses. Therefore, our participants suggested that there needs to be systematic training offered to the academics to clearly understand the importance of the intersectional perspectives in solving real-world cyber security challenges and ways of embedding these concepts in the courses they teach. \blockquote [P2]{We need some culturally sensitive training to help us.
Come up with this sort of material or these sorts of course designs. Because again, you'd be doing it in that sort of isolationist, patronising sort of viewpoint if you weren't aware of each of these different members of society and what their concerns. So timely training}. Our participants further highlighted the importance of providing such training to program managers and discipline leaders to ensure these initiatives are supported at all levels. 

\subsubsection{FW3: Institutional commitment}

\paragraph{FW3.1 University-level integrated approach}
Participants highlighted the need for a university-level integrated approach to better align  Australian HE cyber security curriculams with intersectional perspectives. This should be demonstrated through consistent top-down leadership and strategic allocation of resources.  University KPIs should be re-evaluated to assess their alignment with intersectional goals and initiatives. By ensuring KPIs encompass intersectional perspectives, program managers and discipline heads are motivated and made responsible for driving meaningful initiatives that promote diversity and inclusivity aspects within their curriculams  \blockquote[P4]{This needs to be a top priority in cyber because otherwise, we're really missing all the quality in the market. So maybe developing Key Performance Indicators (KPI) for managers and for the university  to measure how inclusive our curriculum is}.

Once university-level KPI's are in place for recognising the importance of intersectional aspects, there will be motivation to introduce program-level objectives and course learning outcomes that explicitly incorporate intersectional perspectives as well.  Such integration helps to create a curriculum that not only meets academic standards but also prepares students to engage effectively with complex, real-world challenges in a diverse society. Furthermore, participants emphasised the importance of aligning classroom discussions with student assessments. It's crucial to focus on developing assessments that incorporate elements of intersectionality, ensuring that diverse perspectives are reflected and understood by the students  \blockquote [P7]{If I talk about the attacks on under-represented groups, then there should be good questions in my assignments or exams using real-word scenarios}.

\paragraph{FW3.2 Introduction of elective courses}
To ensure that Australian higher education is inclusive, relevant, and responsive to the intersectional needs of diverse student populations, universities could begin by introducing related elective courses to their curriculams rather than introducing the intersectional perspectives in core courses directly. This may be viewed as a more risk-averse approach as the integration can be accomplished without significantly altering existing frameworks and structures. The success of such elective courses can serve as a starting point to demonstrate the importance of the associated content to universities, government, industry, and accreditation bodies \blockquote [P9]{I actually think what you would need to do is develop a unit as an elective fast into a degree program.  To then highlight the success of that new offering, engage it with the industry.  Then, suggest developing it into the core of the degree course as part of that conversation, having the engagement.  With government and industry as part of that, the elective can develop and grow. 
The problem you're going to have is taking something that's a new concept and embedding it directly into the core of a degree course.  When it comes to the curriculum in universities, it's very much risk-averse. So again, when you have a new idea or are trying to fit that into the core, you may face quite a bit of resistance. Whereas if you developed the new unit as an elective and really proved the concept, you know, raised the awareness of the elective, then the next step is to get it into the core of the degree}.

A participant also highlighted that program managers and discipline leads should carefully consider who will teach such courses. One participant explained that academics who teach such courses should be carefully selected based on their passion and teaching and research interests, rather than selecting someone from the team who has a lighter teaching load. 

\paragraph{FW3.3 More recognition and support for academics from their departments and disciplines}

As mentioned in Section~\ref{CH:lackofgovernmentsupport}, academics are often overloaded with research and teaching commitments. As a result, even academics motivated to make changes to their courses are unable to do so due to constraints in their workloads. Therefore, our participants noted that discipline heads or school heads must actively seek ways to support such academics by providing them with additional resources or allocations within their workload to develop new courses and/or improve existing ones \blockquote[P13]{The workload model should factor these as well. Because if that is important, then it is not just a matter of ticking the boxes - Oh yeah, we got that interdisciplinary thing, or do we really need to deep dive into it?  Then that should be clearly identified in the workload. It is like developing a new state-of-the-art master's program}.

On the other hand, academics are often driven by career progression and promotions; hence, universities could consider ways to provide more recognition to academics who make an effort to update their courses or develop new courses with intersectionality aspects in mind. Such incentives would encourage more academics to consider these perspectives when designing their courses.

\paragraph{FW3.4: More women, gender diverse and CALD representation in staff}

Participants emphasised the importance of increasing the representation of women, gender-diverse individuals, and those from CALD backgrounds among academic staff, highlighting their potential to serve as role models for students. Moreover, involving such academics in curriculum design can enrich the process by incorporating their lived experiences and diverse perspectives \blockquote [P8]{Something that we did when I was still in [a different University], and it goes back to what I mentioned earlier about modelling and seeing role models, is that we very intentionally recruited female lecturers for some positions to be able to have a role model. After we recruited her, she was a data scientist, we started have a lot more females wanting to participate in that. So I mean something as simple as that. It's just a role model who shows that this is possible can be done. You know you should never disregard how how impactful that can be}.

\subsubsection{FW4: Cross-Sector collaboration and partnerships}

\paragraph{FW4.1: Government support and funding}
 Our participants emphasised the need for the government to play a pivotal role in promoting inclusivity and intersectional aspects within society as a whole. The government could consider targeted funding or incentives for universities that meet or exceed intersectionality goals.   \blockquote [P14]{The government could fund and support some institutions that are offering courses like this. So that the institution itself. Universities aren't the ones that are driving it. It would be more government-based and you could then get the right experts who have the right expertise in the room to develop and deliver those courses}.  One of our participants even highlighted that Australia does not have a dedicated cyber security minister, and the current cyber security minister oversees many sectors where priorities may be different \blockquote[P9]{We had a dedicated Cyber Minister, who sat in Cabinet as a cyber minister. That isn't the case now}. Furthermore, participants noted that the government could encourage more research in this area by enabling anonymised, shared datasets, as is done in healthcare.

\paragraph{FW4.2: Collaboration between academia, industry and government}

Our participants emphasised the importance of fostering stronger partnerships and shared visions among government, industry, and academia to promote and support intersectional perspectives within the University. Such partnerships would then be reflected in top-down leadership, resource allocation, and systemic accountability from all parties, including universities, government and industry.  One participant explained that the recently established executive cyber council in Australia and the cyber security skills workbook that is planned to be launched could be considered the first steps towards that much-needed collaboration in the cyber security sector. However, the participant was not able to comment on both these initiatives further as the information about them was not publicly available at the time of the interview \blockquote [p9]{I suppose there is a need to bring together industry, academia, and government, whether the Executive Cyber Council is the way to do that, I don't know. But it's the first step in that way.  Within the next sort of month or so, this cybersecurity Skills workbook will be released. What's in that?  I do not know.  But because it would be supported by the government and then supported by industry, it'll be used as sort of a major driver of change}.

Furthermore, participants discussed how the industry could contribute through funding, sharing real-world use cases, and providing access to data. One participant highlighted that Work Integrated Learning (WIL) could be an effective mechanism to enhance these collaborations. While some industries have a genuine need to tackle pressing issues, academics can play a crucial role in supporting these efforts. In return, industry partners should be encouraged to actively engage, providing relevant data to foster meaningful collaboration and impact. \blockquote[P11]{Usually what happens is industry says I have a problem and academics solve it for me. Then the academics do what they can, and they come up with a report that industry may or may not then integrate. But if industry actually does stuff, if it actually gets involved in the teaching or in taking on interns or in sharing some potentially useful data that could be useful for students.  To work with or create opportunities for them to role-play or create use cases that we can then use. That would be something that the industry could genuinely do to help and support}.
\section{Threats to validity} \label{threats}
We discuss potential threats to the validity of the study, specifically in terms of data collection and analysis. The data collection materials, designed to capture comprehensive insights, were carefully crafted to minimise biases and maximise participant input. All authors contributed to the development of the interview guide.  Based on the insights gained from the pilot interview, we decided to include a set of slides at the beginning of the interview to provide participants with a clearer understanding of the study's aims. Furthermore, we invited participants to share either their own experiences or those of others whom they know. The research team intentionally selected this approach to cultivate a supportive and non-confrontational atmosphere, ensuring participants felt secure and not scrutinised. By fostering this environment, we aimed to facilitate open and honest dialogue, thereby enriching the quality and depth of the insights gathered. 

Our participants included three Professors, four Associate Professors, three Senior lecturers and five lecturers who had experience in teaching and coordinating postgraduate
and undergraduate cyber security courses.  While the academics who participated in this study represented all states in Australia and had a vast amount of experience in the field, we acknowledge that qualitative studies are inherently interpretive \cite{malterud2016theory}; the findings are based on the studied context, thus making it challenging to generalise. We reached theoretical saturation, which indicates adding more participants from the same group will not influence findings; however, large-scale studies can be conducted to validate the current study’s findings and investigate their generalisability in other  contexts.  We believe our findings can be recreated and adapted in other similar contexts. To support this, we have provided detailed information about the participants, research methodology,  and  the interview material in Section~\ref{studymethods}. 

Data analysis was primarily conducted by the first two authors. Both authors jointly conducted the data analysis through an iterative and collaborative process. They regularly reviewed and discussed the emerging codes, sub-themes, and themes in relation to the interview transcripts to ensure consistency, accuracy, and analytical rigor. All interpretations were continuously cross-validated against the raw data, drawing on the each authors’  expertise of more than 10 years in qualitative research and analysis.  They held meetings to discuss the findings in depth, resolving any disagreements collaboratively. However, the findings were not verified with participants through member checking, posing a risk of potential misinterpretations from the interviews.


\section{Discussion and conclusion} \label{dicussion}
This paper critically examines cybersecurity education within the Australian Higher Education sector, focusing on the inclusion of women, gender-diverse individuals, and those from CALD backgrounds. Our findings reveal  complex  barriers and essential future actions, which ultimately affect Australia's ability to develop a robust, innovative, and diverse cybersecurity workforce  \cite{eSafety2019, APSC2024CALDstrategy}. In this section, we provide an overview of study findings and discuss implications of study findings for universities and policy makers, educator, industry and researchers. 

\subsection{Overview of findings}
 
Overall, our findings indicate that intersectional perspectives remain insufficiently integrated within the Australian HE cybersecurity curriculum  \cite{eSafety2019, APSC2024CALDstrategy}. This observation aligns with prior research by Vykopal et. al., (2025), which demonstrates that many Cyber Security programs globally—despite their labeling—fail to incorporate essential interdisciplinary and non-technical dimensions, highlighting a broader systemic gap between curricular design and contemporary competency expectations \cite{vykopal2025cybersecurity}.
 
Although our participants pointed out some ad-hoc ways to integrate intersectional perspectives into the Australian HE cybersecurity curriculum, there is a lack of systematic ways to embed women, gender-diverse, and CALD perspectives in the Australian HE cybersecurity curriculum \cite{sheanoda2024sexuality, leyton2021culturally, anwar2017gender, namukasa2023understanding, peacock2017gender}. Existing practices are often driven by individual staff interests or one-off projects, rather than embedded through coordinated program design, assessment, or accreditation requirements \cite{abeysooriya2023discussion, mountrouidou2019securing}. As a result, diversity topics are often treated as add-ons and inconsistently implemented, limiting its impact on student learning and on the broader culture of cybersecurity education.

The analysis of barriers to incorporating diverse perspectives in cybersecurity curricula uncovers deeply rooted issues that go beyond mere knowledge gaps and recruitment challenges  \cite{parrish2018global, mountrouidou2019securing}. Participants highlighted structural constraints such as crowded curricula, rigid accreditation standards, limited time and resources for curriculum an course revamping, and a shortage of accessible, context-specific teaching materials. These challenges are also intensified by cultural and other organisational issues \cite{eSafety2019, APSC2024CALDstrategy, HenryFlynn2018, griffin2024sfia}, such as common views of cybersecurity as a highly technical, male-dominated field, some educators’ lack of confidence in teaching intersectional topics, and limited institutional motivation to make inclusivity a priority. Together, these factors create a cycle in which diverse perspectives are viewed as desirable but non-essential, and therefore remain marginal.

Finally, our findings on future work and support needed underscore a critical transition from ad-hoc initiatives to a more strategic, integrated, and empirically driven approach \cite{eSafety2019, APSC2024CALDstrategy, griffin2024cybersecurity}. Participants called for contextual frameworks and guidelines that explicitly articulate how diversity and inclusion can be embedded across core cybersecurity units, alongside repositories of shared teaching resources and case studies that centre women, gender-diverse, and CALD experiences \cite{vykopal2025cybersecurity}. They also emphasised the importance of targeted professional development, dedicated funding, and recognition for staff who lead inclusive curriculum innovation. Sustained collaboration between universities, industry, and community partners will be essential to co-design and evaluate interventions that are  scalable and sustainable and also responsive to the lived experiences of marginalised groups in cybersecurity.

\subsection{Implications for practice}
In this section we discuss the implications of our study findings to universities,  educators, industry and researchers.

\subsubsection{Implications for universities}
\begin{itemize}
  \item \textbf{Curriculum approval and review}
Course and programme approval templates could explicitly require evidence of how intersectional perspectives  are embedded in learning outcomes, content and assessment \cite{vykopal2025cybersecurity, accenture2026cyberworkforce, ACCC2024, eSafety2019, acara_digital_technologies_f10, griffin2024sfia, straight2024beyond}. This moves inclusion from being an optional add‑on which educators often neglect to a core quality criterion. For example, course learning outcomes could require students to critically analyse how cybersecurity threats, policies, and technologies impact different communities, rather than assuming a universal user \cite{vykopal2025cybersecurity, accenture2026cyberworkforce}. Review panels could be asked to comment specifically on whether units demonstrate diverse case studies, and assess students on their ability to recognise and address bias in cybersecurity systems \cite{vykopal2025cybersecurity}.
\item \textbf{Workload and promotion recognition}
Universities should formally recognise inclusive curriculum development as academic work, rather than treating it as voluntary work. Our findings indicate that allocating time in the workload  for the design, piloting and evaluation of intersectional teaching materials and recognising such contributions in promotion, performance review and teaching awards may motivate academics to initiate and enagage in such activities. Without this recognition, inclusive curriculum work will continue to depend on a small number of committed individuals which risks burnout and reinforces existing inequities.

  \item \textbf{Resourced interdisciplinary teams}
  Our findings suggest that meaningful inclusion of women, gender-diverse and CALD perspectives cannot be achieved by cybersecurity academics working in isolation. Universities should invest in interdisciplinary teams that bring together expertise from cybersecurity, education, gender studies, sociology, and cultural studies to co‑design curricula and assessment. Providing  funding for such collaborations, along with administrative and learning design support, would enable the development of richer, contextually grounded materials that address both technical and socio‑cultural dimensions of cybersecurity. 

  \item \textbf{Engagement with accreditation bodies}
  Accreditation and professional bodies have a crucial role in shaping the core cybersecurity knowledge and skills \cite{asd_cyber_skills_framework_2020,griffin2024sfia, straight2024beyond}. Universities and policy‑makers should actively engage these bodies to advocate for standards that explicitly recognise diversity, equity and inclusion competencies as integral to professional practice. Such alignment would ease the perceived conflict between satisfying accreditation requirements and embedding inclusive content, and would clearly signal to providers that intersectional perspectives are fundamental to producing industry‑ready graduates and  not something optional.
\end{itemize}

\subsubsection{Implications for  educators}
\begin{itemize}
  \item \textbf{Integrate intersectionality into all aspects  of teaching and learning}
   Rather than treating diversity as a stand‑alone topic, educators can embed intersectional analysis into existing course content, assignments and exercises.  This might involve asking students to consider how threat models, authentication mechanisms, or security policies affect users differently based on gender, culture, language, disability, or migration status. For example, students could be required to critique a security protocol from the perspective of a multilingual user, or redesign an access control system to better accommodate persona of a  gender‑diverse staff. Embedding these questions within core technical tasks helps students see intersectional thinking as an integral part of rigorous cybersecurity practice, rather than an optional ethical add‑on \cite{vykopal2025cybersecurity}.

  \item \textbf{Co-create content with industry and communities}
  Educators could  seek opportunities to co‑develop teaching materials with industry partners, community organisations, and advocacy groups that work with under‑represented populations \cite{vykopal2025cybersecurity, accenture2026cyberworkforce}. This co‑creation also helps keep examples relevant to current threats and real workplace situations, while reducing the risk of using  stereotypical representations of diversity.

  \item \textbf{Targeted professional learning}
   Many of the participants reported limited confidence in teaching intersectional content. Universities and professional bodies should therefore offer targeted professional learning opportunities—such as workshops and mentoring—that focus on inclusive pedagogies for cybersecurity \cite{vykopal2025cybersecurity, griffin2024sfia,accenture2026cyberworkforce, asd_cyber_skills_framework_2020}. These should provide practical strategies for educators embded intersectional perspectives to their own courses.  Ongoing professional learning, rather than one‑off sessions, will be critical for sustaining change in teaching practice.

\end{itemize}

\subsubsection{Implications for  industry audience}
\begin{itemize}
  \item \textbf{Develop research-informed training and awareness program for their future employees}
Industry partners can utilise research findings to design training and awareness programmes that equip graduates to recognise, critically examine and address diversity and inclusion issues in cybersecurity practices within their workplaces \cite{vykopal2025cybersecurity, griffin2024sfia,accenture2026cyberworkforce, asd_cyber_skills_framework_2020}. This content should be embedded in the onboarding process and ongoing professional development, so that inclusive cybersecurity practices are reinforced throughout employees’ careers and contribute to lasting shifts in workplace culture.
  \item \textbf{Develop research-informed apprentices program}
  Apprenticeship and early‑career pathways can be deliberately structured to support diverse entrants to cybersecurity \cite{vykopal2025cybersecurity, griffin2024sfia,accenture2026cyberworkforce, asd_cyber_skills_framework_2020}. Industry can work with researchers and educators to identify barriers faced by women, gender‑diverse and CALD candidates, and to design supports such as mentoring and targetted learning.
\end{itemize}

\subsection{Implications for research}
\begin{itemize}
  \item \textbf{Including student and community perspectives}
 Future research should more systematically centre the perspectives of students and community stakeholders \cite{vykopal2025cybersecurity, griffin2024sfia, ACCC2024}. This includes investigating the cybersecurity experiences, barriers and learning needs of marginalised groups; how they engage with and experience cybersecurity curricula and learning environments; what motivates or discourages them from persisting in the field. We also anticipate a stronger emphasis on co‑design, where students and community partners work alongside educators and researchers to shape pedagogical approaches and curriculum content.

  \item \textbf{Contextualised frameworks and toolkits}
  Our findings provide evidence that there is a need to develop and test frameworks and practical toolkits that help educators integrate intersectional perspectives into cybersecurity in ways that are sensitive to local context \cite{vykopal2025cybersecurity}. This might involve adapting existing international models of inclusive STEM education to the specific cultural and institutional context of Australian higher education, and developing practical resources such as curriculum‑mapping templates, sample rubrics, and checklists to support inclusive cirriculam and course  design.

  \item \textbf{Evaluating intersectional pedagogical interventions}
  Finally, empirical evaluation of intersectional pedagogical interventions in cybersecurity is needed \cite{vykopal2025cybersecurity, griffin2024sfia}.  For example, future studies could investigate the impact of different teaching strategies (e.g. scenario‑based learning, community‑engaged projects, reflective journals) on student learning outcomes, attitudes towards diversity. Longitudinal research can track whether exposure to intersectional curricula influences graduates’ professional practices and their contributions to workplace culture and policy. Such evidence will be critical for refining interventions, justifying institutional investment, and influencing accreditation and policy settings.
\end{itemize}

In summary, this study shows that intersectional perspectives are still marginal within Australian higher education cybersecurity curricula, despite growing recognition of their importance. Systemic constraints—spanning curriculum structures, accreditation requirements, institutional cultures, and limited staff support—mean that diversity and inclusion are often treated as optional rather than integral to professional competence. Addressing these challenges will require coordinated action by universities, educators, industry and researchers to embed intersectionality into curriculum design, accreditation standards, professional learning and workplace practices \cite{vykopal2025cybersecurity, griffin2024sfia}. Doing so is essential not only for equity, but for building a more resilient, innovative and socially responsive cybersecurity workforce for Australia.

\begin{acks}
We sincerely thank all the participants who generously shared their time and experiences during the interviews with us. The work has been supported by the [removed for anonymity]. The authors would also like to acknowledge the use of AI tools like Grammarly, Microsoft CoPilot, and
ChatGPT to assist with proofreading and improve the clarity of some authors’ own text in the manuscript.

\bibliographystyle{ACM-Reference-Format}
\bibliography{sample-base}

\end{acks}
\end{document}